\documentclass[prc,twocolumn,showpacs]{revtex4-1}
\usepackage{graphicx,amsmath,amssymb,bm}

\newcommand{\bra}{\big\langle}
\newcommand{\ket}{\big\rangle}
\newcommand{\be}{\begin{equation}}
\newcommand{\ee}{\end{equation}}
\newcommand{\mev}{\, \text{MeV}}
\newcommand{\gev}{\, \text{GeV}}

\newcommand{\ceft }{\,\chi \text{EFT}}

\begin{document}

\title{Correlations in light nuclei and their relation to fine tuning and uncertainty quantifications of many body forces in low-energy nuclear physics}

\author{S.\ Lupu}
\affiliation{Racah Institute of Physics, 
The Hebrew University, 
91904 Jerusalem, Israel}
\author{N.\ Barnea}
\affiliation{Racah Institute of Physics, 
The Hebrew University, 
91904 Jerusalem, Israel}
\author{D.\ Gazit}
\email[E-mail:~]{doron.gazit@mail.huji.ac.il}
\affiliation{Racah Institute of Physics, 
The Hebrew University, 
91904 Jerusalem, Israel}

\begin{abstract}
The large nucleon-nucleon scattering length, and the isospin approximate symmetry, are low energy properties of quantum chromodynamics (QCD). These entail correlations in the binding energies of light nuclei, e.g., the $A=3$ iso-multiplet, and Tjon's correlation between the binding energy of three and four body nuclei. Using a new representation of these, we establish that they translate into a correlation between different short-range contributions to three body forces in chiral effective field theory of low-energy nuclear physics. We demonstrate that these correlations should be taken into account in order to avoid fine-tuning in the calibration of three body forces. We relate this to the role of correlations in uncertainty quantification of non-renormalizable effective field theories of the nuclear regime. In addition, we show that correlations can be useful in assessing the importance of forces induced by renormalization group (RG) transformations. We give numerical evidence that such RG transformations can be represented effectively by adding a constant to the pure three nucleon contact low energy constant $c_E$. 
\end{abstract}


\maketitle
\section{Introduction}
Nuclear physics is a low-energy realization of Quantum Chromodynamics (QCD). However, the latter is 
non-perturbative at these low-energies, leaving this fundamental connection hidden behind seemingly unrelated
properties.

The renormalization group (RG) and effective field theory (EFT)  \cite{1979PhyA...96..327W,*1990PhLB..251..288W,*1991NuPhB.363....3W,*1984AnPhy.158..142G} approaches have been used for a quadranscentennial to give new insights to this problem, opting on the separation of scales in the nuclear sector. The fact that the nucleon-nucleon scattering lengths are large compared to the nuclear force range, as 
indicated by the unnaturally small deuteron binding energy, has led to the development of ``pionless'' EFT, a renormalizable theory of point SU$(4)$ nucleons interacting via contact interactions \cite{Kaplan1997471}. As only the coulomb force breaks neutron-proton symmetry at leading order (LO), binding energies of iso-multiplets are correlated, and in particular of $^3$H and $^3$He \cite{PhysRevC.67.034004,0954-3899-37-10-105108,PhysRevC.83.064001,PhysRevC.89.064003,0954-3899-42-4-045101}. For $A=3$, pionless EFT has a limit cycle behavior  \cite{1970PhLB...33..563E,*PhysRevLett.82.463}, which shows the need of introducing a three nucleon force (3NF) already at LO. A single 3NF parameter naturally creates a correlation between three nucleon observables, such as the neutron-deutron doublet scattering length and the $A=3$ binding energy, i.e., recovers the 
previously unexplained empirical correlation also known as the ``Phillips line'' \cite{Phillips1968209,Platter2005254,EPJA.44.1434}. 
Once such 3NF is introduced at LO, no significant cutoff variation is found in the binding energy of the $A=4$ system \cite{Platter2005254}, 
implying that 4NF is unnecessary at this order, and creating a correlation between the $A=3$ and $A=4$ binding 
energies, i.e. the ``Tjon line'' \cite{Tjon1976391,Platter2005254,EPJA.44.1434}. 
Pionless approach is only applicable for low momentum phenomena, and as a result 
is limited to light nuclei, where the binding energy $B$ is small enough, such that typical nucleon momentum $Q_B$ is much smaller than the pion mass $m_\pi$, $Q_B\approx \sqrt{M_NB/A}\ll m_\pi$ ($M_N$ is the nucleon mass). For heavier nuclei $Q_B \sim m_\pi$, and thus the theory needs to include pions as viable degrees of freedom. 

This is accomplished via Chiral EFT ($\chi$EFT) \cite{1979PhyA...96..327W,*1984AnPhy.158..142G,*1990PhLB..251..288W,*1991NuPhB.363....3W}. $\chi$EFT follows the notion that properties of the strong force are linked to the approximate chiral symmetry of QCD, broken at low energy to the isospin symmetry $\text{SU}(2)_R\times\text{SU}(2)_L \rightarrow \text{SU}(2)_V$. The pion is identified as Nambu-Goldstone boson of this symmetry breaking, and the pion decay constant  $f_\pi=92.4\mev$ is the order parameter. This indicates that the chiral symmetry breaking scale is $\Lambda_\chi \approx 4\pi f_\pi \approx 1\gev$. This inherent scale separation, $Q_B\sim m_\pi \ll \Lambda_\chi$, invites the introduction of $\chi$EFT for the description of the nuclear regime. $\chi$EFT is constructed by integrating out chiral scale degrees of freedom, leaving only nucleons and pions as explicit degrees of freedom \footnote{We note that with this choice of degrees of freedom, the actual breakdown scale is significantly lower than $1 \gev$, i.e., $\Lambda\sim 500 \mev$}. The resulting effective Lagrangian retains all symmetries, particularly the approximate chiral symmetry, of the underlying theory. Weinberg additionally showed that it can be organized in terms of a perturbative expansion in positive powers of $Q/\Lambda_\chi$. The coefficients of the different terms in the Lagrangian are called low energy constants (LECs). They capture the integrated out physics, and in general depend upon the cutoff. Currently, LECs are fixed to reproduce experimental data. 

Since its inception, chiral EFT, has been used to derive nuclear forces \cite{RevModPhys.81.1773}, as well as nuclear N\"other currents induced by the global symmetries of the chiral Lagrangian \cite{2003PhRvC..67e5206P,*Park1993341,*1996NuPhA.596..515P,*1994NuPhA.579..381P,*2008PhDT.......216G,*Bacca2014}. Though the non-perturbative extension of the expansion has been shown to have some weaknesses, that induced a debate on the correct ordering of the expansion \cite{PhysRevC.72.054006,PhysRevC.74.014003}, high order $\ceft$ at relatively high order shows good success in predicting nuclear properties. Systematic order-by-order improvement is needed to be demonstrated in order to assess the EFT propeties of $\ceft$, especially for nuclei heavier than $A=3$. 
In the way $\ceft$ is usually practiced, i.e., Weinberg power counting, where the expansion is about a trivial zero momentum fixed point, implying na{\"i}ve mass dimension analysis, the theory is non-renormalizeable, and thus cutoff variation is limited \cite{PhysRevC.72.054006}. 

One of the implications is that assessing the accuracy of predictions, as well as quantification of uncertainties, which in EFTs rely sometimes on cutoff variation, are challenging. 
Moreover, in the fitting process of LECs to observables, EFT inherent theoretical truncation error is combined with experimental errors \footnote{Note that most potentials fit to phase-shifts, where no experimental errors are assigned. Ref.~\cite{2015arXiv150602466C} uniquely deals with this problem} . Enormous progress has been made in the last couple of years in assessing errors in predictions of EFT \cite{PhysRevLett.110.192502,2015arXiv150602466C,2015JPhG...42c4028F,2015arXiv150601343F,2014arXiv1412.4623E}. While applying very different approaches, all these references cope with the problem of how to estimate the combined error arising from EFT truncation and simultaneous fit of dozens of LECs such that the model will reproduce the enormous amount of data from nucleon-nucleon scattering. In general, order-by-order analysis is performed, accompanied by clever ways to propagate experimental error. In addition, it seems that dictating ``naturalness'' of the LECs, e.g., by using a Bayesian prior, is a very efficient way to get the expected EFT order-by-order convergence \cite{2015JPhG...42c4028F,2015arXiv150601343F}. 

In the current paper we concentrate on the calibration process of three nucleon forces within chiral EFT. This poses a different challenge, due to the small number of 3NF LECs. Up to fourth order (N$^3$LO) there are only two LECs that appear exclusively in 3NF. We achieve this by employing known LO  correlations in pionless EFT, in particular the Tjon's correlation and isospin invariance of the nuclear forces. Not only does this allow to identify mistakes in the calibration of many body forces, but also to assess higher order contributions to observables and to avoid fine tuning in LECs calibrations. These make correlations whose origin is a different EFT to the same theory a powerful tool in the standing problem of quantifying the uncertainty in theoretical predictions of the EFT. Finally, by observing the evolution of such correlations to lower cutoffs using renormalization group transformations, we conjecture that the breaking of such correlations is a signature of neglected terms in the EFT potential.

\section{Calibration of $\ceft$ forces}

\begin{figure}[t]
\begin{center}
\rotatebox{0}{\resizebox{5.5cm}{!}{
\includegraphics[clip=true,viewport=2.5cm 18.5cm 20.5cm 27.5cm]{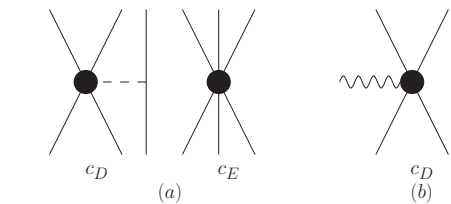}} }
\end{center}
\vspace*{-5mm}
\caption{Contact and one-pion exchange plus contact interaction (a), and axial current contact (b) terms of $\ceft$.}
\label{currents}
\end{figure}

The nuclear potential derived in $\ceft$ is systematically ordered \cite{VanKolck1999337,PhysRevC.53.2086,Epelbaum2000295,Epelbaoum1998107,PhysRevC.68.041001,Machleidt20111,PhysRevC.90.054323}. For example, the leading order (LO) includes 
a pion exchange force, and
singlet and triplet two nucleon contact terms. The next-to leading order (NLO) contribution includes corrections to the two body nuclear potential suppressed by $\left(\frac{Q}{\Lambda}\right)^2$ with respect to the LO. $\left(\frac{Q}{\Lambda}\right)^3$ suppressed terms appear at N$^2$LO, in which first 3NF appear \cite{PhysRevC.49.2932,PhysRevC.66.064001,PhysRevC.68.041001,Machleidt20111,2007FBS....41..117N}.
$\ceft$ at N$^2$LO includes three different 3NF topologies: (i) N-$\pi$-N-$\pi$-N, i.e., two pion exchange; (ii) N-$\pi$-NN, i.e., two-nucleon contact coupled to an additional nucleon via pion exchange; and (iii) NNN, i.e., three nucleon contact. The two latter, (see Fig.~\ref{currents}(a)) include two new low-energy constants (LECs) that do not appear in the two nucleon or pion-nucleon sector, coined $c_D$ and $c_E$, respectively, while the long range two pion exchange term depends on pion-nucleon scattering parameters, that appear also at the two body level of N$^2$LO $\ceft$, $c_1$, $c_3$, $c_4$. It is important to note that at N$^3$LO no new 3NF LECs appear. Up to N$^2$LO there are about 20 LECs that appear in the nuclear forces. The calibration of $\ceft$ forces represents a multi-variable optimization problem \cite{PhysRevLett.110.192502}

In Ref.~\cite{2015arXiv150602466C}, the authors test a simultaneous fitting approach. This is in contrast to the separate, sequential approach. In the latter, one notes that the LECs can be divided into two main sections, i.e., contact interactions between nucleons which represent integration out of physics above the chiral symmetry breaking scale, and $\pi$-N LECs  which are of longer wavelength. The sequential approach fixes these groups separately. This is due to the different momentum sensitivity (UV vs. lower energies), as well as the different physical origin, i.e., $\pi$-N LECs can be calibrated using pion-nucleon scattering observables only, whereas NN LECs have to include NN scattering data to be calibrated. This kind of separation has motivated bayesian approach \cite{2015JPhG...42c4028F,2015arXiv150601343F} to separate short-range and long-range contributions to the calculation error. Moreover, different locality properties of the forces can be demonstrated by testing the convergence pattern when applying different regulators for these terms \cite{doi:10.1146/annurev-nucl-102010-130056}. 

Ref.~\cite{2015arXiv150602466C} has tried both. The simultaneous fixing is shown to have smaller order-by-order uncertainty evolution, and better predictivity. By carefully studying the optimization, the authors show major differences between the approaches. In particular, they show that the covariance matrix of the sequential approach is nearly diagonal, while strong cross-correlations appear in the simultaneous approach. This is related to the fact that the experimental data of the $\pi$-N sector is not constraining well enough the theory, due to too large experimental errors. In addition, it is clear that one cannot separate the $\pi$-N effects in the NN sector, thus the latter can be used also to constrain the former. However, all these observations can also hint to possible fine tuning. EFT is susceptible to fine tuning due to the fact that it is a truncated expansion, and thus has an inherent error. In the following we focus on the role of correlations in fine tuning of 3NF LECs. 

Sequential fixing of 3NF LECs has been accomplished by the use of different observables. The triton b.e., together with the neutron-deuteron doublet scattering length \cite{PhysRevC.66.064001}, or with $^4$He b.e. \cite{PhysRevC.70.061002,*PhysRevC.73.064002}. Heavy nuclei properties were also used \cite{PhysRevLett.99.042501,*PhysRevC.91.051301}.  
Lately, the fact that $c_D$ appears in the weak interaction operator was used to calibrate this 3NF LEC, using the triton b.e. and decay rate as two constraints \cite{PhysRevLett.103.102502,PhysRevLett.96.232301}. Albeit the half-life is not as accurately measured as the aforementioned observables, this procedure has led to unprecedented accuracy of $c_D$ and $c_E$ fixing, a fact which we argue is related to the fact that the strong observables are correlated.  

The origin of the connection between the weak interaction and 3NF is $\text{SU}(2)_R\times\text{SU}(2)_L$ symmetry, which is common to the chiral symmetry and the weak gauging. In the strong sector, one uses the chiral symmetry of the $\ceft$ Lagrangian to derive its N{\"o}ther currents \cite{1994NuPhA.579..381P,1996NuPhA.596..515P}. These too are ordered according to their relative importance, as LO, NLO ($\left(\frac{Q}{\Lambda}\right)$ suppressed), etc. As this is exactly the weak interaction gauging, these currents couple to weak interaction probes. In particular, weak reaction rates depend, up to kinematical factors, upon matrix elements of these currents between the initial and final nuclear states. Thus, these currents, together with wave functions can be used to calculate, directly from $\ceft$ weak processes. It is instructive to understand the relevant structure of these currents in the nucleus. Up to $\left(\frac{Q}{\Lambda}\right)^2$ (for currents this is N$^2$LO), an axial probe interacts with a single nucleon within the nucleus. However, at $\left(\frac{Q}{\Lambda}\right)^3$ (N$^3$LO), two new topologies appear: an interaction with a nucleon with a simultaneous pion exchange with a different nucleon, and a topology in which the axial probe interacts with two nucleons at contact (see Fig.~\ref{currents}(b)). The latter is derived from the same term in the Lagrangian as the $c_D$ three nucleon force, and is thus 
related to $c_D$ through the expression \cite{PhysRevLett.103.102502},
%\begin{equation}
$\hat{d}_R \equiv \frac{M_N}{\Lambda_\chi g_A} c_D +\frac{1}{3} M_N
(c_3 + 2c_4) +\frac{1}{6}$.
%\end{equation}
Here, $c_3$ and $c_4$ are LECs of the dimension-two $\pi$-N Lagrangian. This relation allows the use of weak observables which are sensitive to the axial current, for the calibration of $c_D$. In particular, the decay rate of triton has been used to extract the axial electric $J=1$ multipole $|\bra ^3 {\text H} \| E_1^A \| ^3 {\text {He}} \ket| = 0.6848 (11)$ transition matrix element between $^3$H and $^3$He wave functions, which are calculated consistently in $\ceft$ \cite{PhysRevLett.103.102502}. In the following we use this method, which has proven successful in many recent calculations (e.g., \cite{PhysRevLett.113.262504,PhysRevLett.110.192503,PhysRevLett.108.052502}).

\section{Correlations in $\ceft$ calculation of $A=3,\,4$ bound states}
The fact that $\ceft$ and pionless EFT are EFTs of the same fundamental theory, means that their predictions should coincide at the relevant energy regimes, in particular Tjon's correlation and $A=3$ iso-multiplet correlation. The two correlations are not found in nature, but in computational realizations of these bound states using different potentials. In principle, they should be revealed by changing the EFT cutoff, similar to pionless EFT.
The limited cutoff variation of $\ceft$ potentials, make 
it impossible to check whether limit cycle behaviour exists in the $A=3$ binding energies, and the relevance of the expansion about a trivial fixed point.
However, systematically reducing the cutoff is possible using RG transformation \cite{Bogner2003265,*Bogner20031,*PhysRevC.75.061001}. Using such a method, it has been shown that the nuclear forces actually exhibits a Tjon line \cite{PhysRevC.70.061002}. In particular, using two body forces only, all reproducing the nucleon-nucleon scattering data, leads to a scatter plot along a Tjon line, correlating the trinuclei and alpha binding energies (BEs), and showing that neglecting three-body forces leads to at  least a 20\% error in reproducing these nuclei BEs. In principle, a different way to generate a Tjon line is by creating different parameterizations of the EFT potentials at a fixed order and regulator, in the spirit of Ref.~\cite{PhysRevLett.110.192502,2015arXiv150602466C}. In any case,
 $\ceft$ leading 3NF should choose the correct BEs on this Tjon Line, up to higher orders.   

The Tjon line is usually plotted as a correlation line between $^3$H and $^4$He BEs. In the following we show that for a specific choice of the two-nucleon potential, $A=3,\,4$ correlations can be represented as a direct correlation between the three body short range LECs $c_D$ and $c_E$. 

The BE of the $^3$H depends on the 3NF. We thus, for a specific choice of two body parameters, scan the ($c_D$, $c_E$) plane, and search for points which reproduce the $^3$H binding energy. As seen in Fig.~\ref{Fig:Bare_potentials}, this results in a line in this plane \footnote{The fact that this produces a line, rather than a scatter of points, suggests some linear response of the observable to changes in the force. We do not study this here but hint that this might be due the fact that 3NF are basically perturbative.}. We repeat this process for $^3$He and find a different line, which follows the trend of the $^3$H line, though not overlapping. 
This means that while $^3$H and $^3$He BEs are correlated, as expected from pionless EFT, they cannot be described simultaneously by $\ceft$. When using EFT at a specific order $\nu$, predictions have inherent inaccuracy, of the order of $\left(\frac{Q}{\Lambda}\right)^\nu$. Thus one interprets the difference between triton and $^3$He as higher order effects. The difference between the lines can be used to asses the EFT systematic uncertainty in fixing these LECs. 

Indeed, this understanding is backed by studies of the origin of the $^3$H and $^3$He BE difference, e.g., Ref.~\cite{PhysRevC.67.034004}, where 85\% of this difference is found to be a result of the coulomb repulsion between protons in $^3$He, which is described at this order of $\ceft$. The rest of the difference is found to be a result of isospin and charge symmetry breaking effects, specifically the proton-neutron mass difference and the difference between neutron-neutron and proton-proton scattering lengths, which are of higher order in $\ceft$. 

We now repeat the same process for $^4$He, i.e., scan the plane ($c_D$, $c_E$) and plot a line that reproduces $^4$He BE. In Fig.~\ref{Fig:Bare_potentials}, we do this for three specific choices of two body force. For all, the difference between this line is of the order of the difference between the $^3$H-$^3$He lines. As the latter is a higher order effect, we conclude that: (i) $^4$He BE is correlated to the trinuclei BE, stemming {\it deus ex machina} from the $\ceft$ formalism and parameter space; (ii) The difference between the experimental $^4$He BE and the one predicted by a force reproducing trinuclei BE is a higher order effect, i.e., it is suppressed by the EFT expansion parameter, which in this case corresponds to the ratio of the effective range to the scattering length.

As an example of an observable which is not correlated, we plot in the $(c_D,c_E)$ plane the region which recovers the $E_1^A$ strength of the triton beta decay. 
As can be seen in Fig.\ref{Fig:Bare_potentials}, in the ($c_D$, $c_E$) plane such a constraint is orthogonal to the strong BEs constraint. 
We augment the line representing the constraint by the experimental error. Other sources of error, such as cutoff dependence of the current operator \cite{Schwenk2015}, are not included here. The cutoff function and the cutoff value we use for the current is identical to that of the NN-$\pi$-N 3NF diagram. 

\begin{widetext}
\begin{center}
\begin{figure}[t]
\includegraphics[width=\textwidth, trim= 0cm 12cm 0cm 0cm]{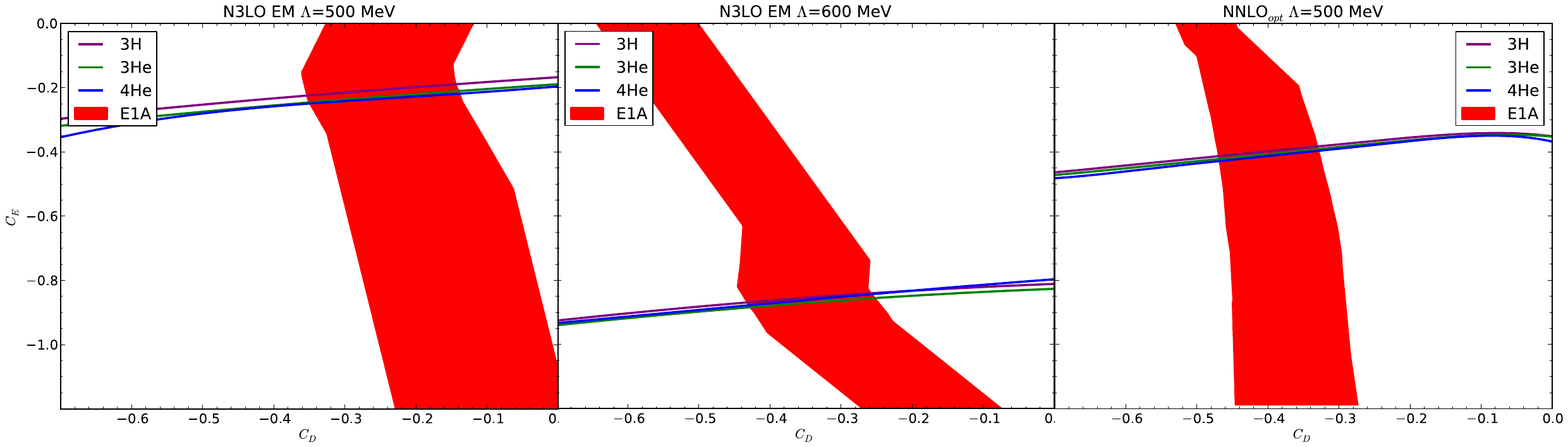}
\caption{Lines in $(c_D,c_E)$ plane which reproduce $A=3,\,4$ bound states, and triton $E_1^A$ strength. $E_1^A$ line is augmented to represent experimental triton half life uncertainty (uncertainty in BEs is smaller than the line width). The lines are plotted with nucleon-nucleon potentials from Ref.~\cite{Machleidt20111} (left pane $\Lambda=500\mev$, middle pane $\Lambda=600\mev$), and from Ref.~\cite{PhysRevLett.110.192502} (right pane).
 \label{Fig:Bare_potentials}} 
 \end{figure}          
\end{center}
\end{widetext}

\section{The Effect of RG transformations on the Tjon Line}

The use of RG transformed potentials, in which momenta lower than a cutoff scale $\Lambda$ are decoupled from the higher ones using exact RG transformations, has become common in recent years \cite{PhysRevC.70.061002,PhysRevC.75.061001,Bogner2003265,Bogner20031,0034-4885-76-12-126301}. These potentials do not include hard core, and thus lead to improved numerical convergence, and allow the use of perturbation theory, while exactly preserving the values of low-energy observables calculated with the original potential. As such, these potentials are ideal for many body calculations. However, such a transformation induces spurious many body forces. In practice, one usually neglects these many body forces at some cluster size (usually clusters above A=3-4 are neglected). 

Assessing the effect of neglecting such cluster forces can be problematic. Here we show that prior knowledge regarding correlations can be used to do so. We do this by using a $v_{low\,k}$ type of potential \cite{Bogner2003265,Bogner20031}, which is constructed using exact RG transformation with a momentum cutoff $\Lambda$, to evolve a microscopic 2-body potential optimized to fit nucleon-nucleon scattering data. In this study we used $v_{low\,k}$ based on the $500\mev$ EM potential \cite{PhysRevC.68.041001}. In Fig.~\ref{Fig:Vlowk_potentials} the correlations representation discussed in the previous section is given for a set of $v_{low\,k}$ evolved to differing momenta. One can observe that as the cutoff is reduced, Tjon's correlation breaking grows. It is interesting to note that the correlation between the $A=3$ nuclei is not broken, a fact that might suggest that it holds to higher orders. Both observations are consistent with a need to include three body forces induced by the RG transformation. 

An interesting effect we find is that Tjon's correlation is broken in a way that can be fixed by adding a constant to $c_E$ LEC. As $c_E$ represents the short range NNN vertex strength, it is reasonable that it would be affected the most by RG transformations, which are constructed to affect only short distances (similar conclusions are known in the NN sector, e.g., Ref.~\cite{2013arXiv1309.5771F}).
Whether a constant shift can be used as an effective way to avoid the technical complication of including 3NF induced by RG transformation 
is a matter of future research, which should be accomplished by observing the behavior of different observables. We note that, phenomenological evidence for the importance of $c_E$ for bulk nuclear properties was also found in Ref.~\cite{PhysRevC.82.024319}.

In Fig.~\ref{Fig:Vlowk_potentials} we plot also the region in $(c_D,c_E)$ plane that recovers the $E_1^A$ strength of the triton beta decay, with use $c_3$ and $c_4$ from Ref.~\cite{PhysRevC.67.044001}. We note, that in principle one should evolve the current using the same RG transformation. Thus, we can give only the qualitative observation, that this observable is indeed useful as an orthogonal constraint for 3NF LECs fixing also for $v_{low\,k}$ potentials. We notice that this constraint gives $c_D$ values  stabilize for small cutoffs values. This is again consistent with RG philosophy of leaving long range observables intact. 

\begin{widetext}
\begin{center}
\begin{figure}[t]
\includegraphics[width=\textwidth, trim= 0cm 5cm 0cm 0cm]{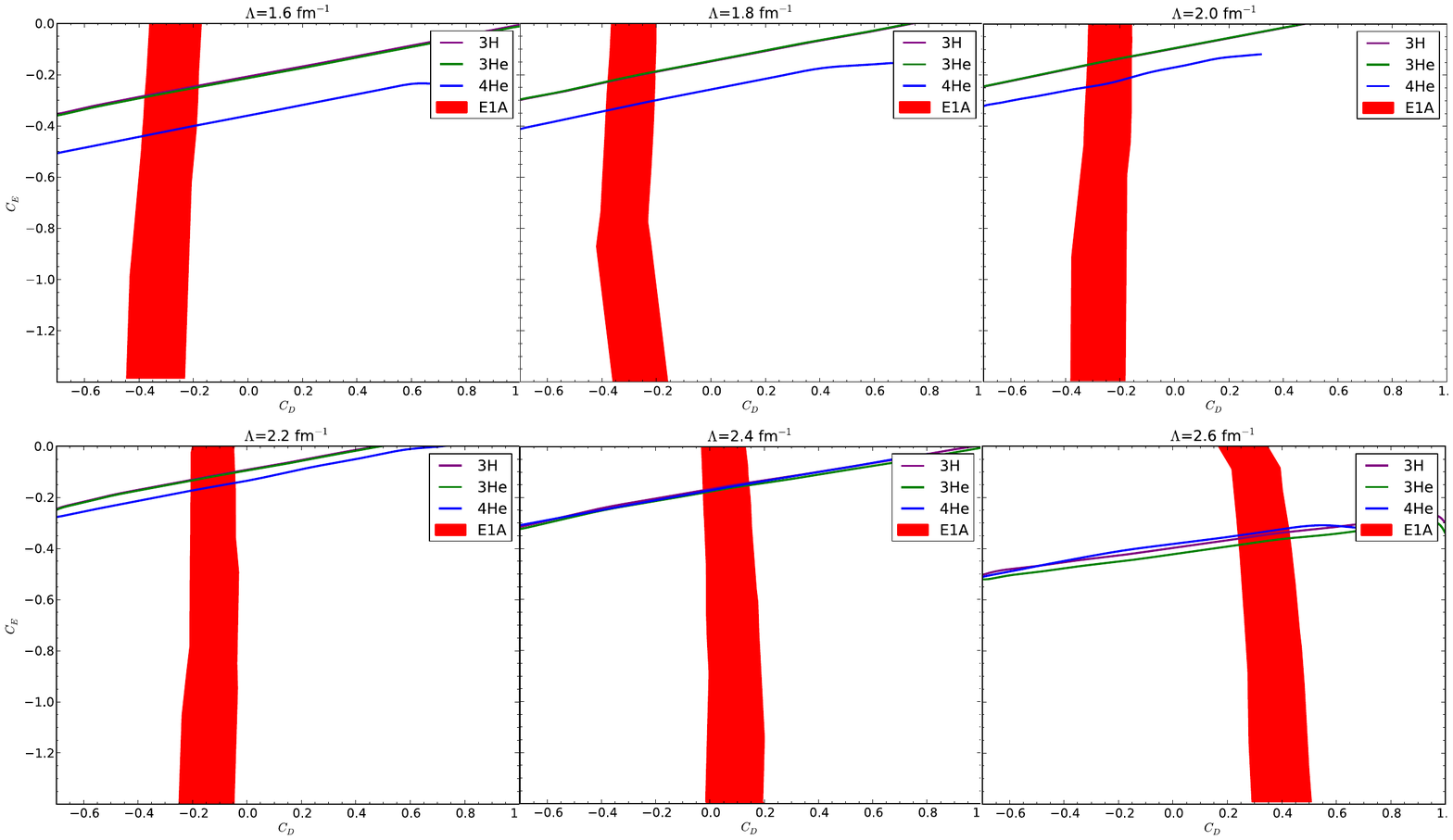}
\caption{Lines in $(c_D,c_E)$ plane which reproduce $A=3,\,4$ bound states, and triton $E_1^A$ strength. $E_1^A$ line is augmented to represent experimental triton half life uncertainty (uncertainty in BEs is smaller than the line width). Lines are plotted with $v_{low\,k}$ flow of of the nucleon-nucleon potential from Ref.~\cite{PhysRevC.68.041001} cutoff at $\Lambda=1.6,1.8, 2.0, 2.2, 2.4, 2.6\,\, {\text{fm}}^{-1}$. 
 \label{Fig:Vlowk_potentials}} 
 \end{figure}          
\end{center}
\end{widetext}

\section{Summary}

Pionless EFT and $\ceft$ correspond to the same fundamental theory, 
and should therefore coincide at very low energies. 
This observation allows utilizing predictions of one EFT as constraints to the other. In particular, pionless EFT predicts at LO correlation between the binding energies of $^3$H and $^3$He, and between $^3$H and $^4$He. We used these correlations to give a prescription for assessing the quality of calibration of $\ceft$ 3NF parameters $(c_D,c_E)$, for a specific choice of two-body nucleon-nucleon potential. Given an observable that depends on 3NF, one plots the region in the $(c_D,c_E)$ plane that recovers this parameter. Using this method for the $A=3,\,4$ binding energies, one finds that they are correlated, though the lines reproducing them are not overlapping. Thus, they are all correlated, as expected by pionless EFT, and the difference between them is interpreted as higher order corrections.

This is in fact a new representation of the correlation between binding energies of light nuclei, including Tjon's correlation. Usually, these were demonstrated by fictitiously changing $^3$H binding energy and showing that the resulting $^{3,\,4}$He BEs lie on top of lines. The current presentation uses physical forces, as they all reproduce real world observables. In fact, had these were the only observables, any point on these lines would correspond to a legitimate nuclear hamiltonian. As such, predictions of other observables using the three different lines reproducing the trinuclei and alpha BEs can be used to estimate EFT truncation errors. The BEs are well known experimentally, to much higher accuracy than that of the EFT estimated this way. Moreover, crossings of the correlation lines, which represent Hamiltonians that reproduce accurately two of these binding energies are coincidental, and might lead to fine tuning. Such fine tuning can result in wrong predictions and underestimation of errors. Contrary to the correlated BEs, we showed that the use of uncorrelated observable, e.g., triton $\beta$-decay rate, is useful as an additional constraint. 
In that case we expect the statistical, experimental, error rather
than the systematical, theoretical, error to be dominant. 

We thus provide a way to disentangle systematic EFT errors from the experimental statistical errors in the observables used to calibrate the many-body forces. Moreover, in the current work we demonstrate that the na{\"i}ve combination of the these two error sources, statistical and systematic, can lead to {\it fine tuning} in the parameterization of the nuclear Lagrangian. Such disentanglement is studied in literature, e.g., using Bayesian approach, simultaneous fitting or order-by-order study, though usually in the context of the problem of calibrating many LECs to many data points \cite{PhysRevLett.110.192502,2015arXiv150602466C,2015JPhG...42c4028F,2015arXiv150601343F,2014arXiv1412.4623E}. Here we showed that in the scenario of a small number of parameters, a fact that usually hints to correlations, a safe calibration method is sequential. It will be interesting to check this method in similar problems, like the neutron-neutron low energy scattering, $M1$ strength, neutron-deuteron scattering length, or pion production \cite{PhysRevLett.96.232301,PhysRevC.90.054001,*PhysRevLett.85.2905,*PhysRevC.77.054001,*PhysRevC.79.064002}. 

In addition, we demonstrated that breaking of these correlations can be used as an indication to missing terms in the calculation. In the case of $v_{low\,k}$ evolved potentials, we showed that Tjon's correlation worsens as the cutoff is lowered. We thus conjectured that this difference is due to neglected induced many body forces in an RG transformation of the potential. Moreover, our results indicate that the majority of the effect of induced many body forces might be effectively represented by a constant (cutoff dependent) shift in the short range 3NF LEC $c_E$. It also seems that there is only a moderate RG flow of the bare weak transition operator. In future work, where other observables will be studied in parallel, we intend to check this. A positive confirmation of these observations might constitute a simple and effective solution to the technical problem of including currents and forces induced by RG transformations, as well as to better understanding regularization and power counting issues.

\begin{acknowledgments}
We thank Sonia Bacca for useful discussions regarding the implementation of $v_{low\,k}$ potential, and Dick Furnstahl, Daniel Phillips, Christian Forss\'en and Andreas Ekstr\"om for the careful reading of the manuscript, as well as valuable comments. 
D.\ G. and S.\ L. are supported, in part, by BMBF ARCHES Award. N.\ B. is supported, in part, by the Israel Science Foundation under grant number 954/09. 
\end{acknowledgments}

\end{document}